\documentclass[journal]{IEEEtran} 				% For final IEEE paper version
\IEEEoverridecommandlockouts 						% For final IEEE paper version
\usepackage{amsmath,amssymb}
\usepackage{amsthm}
\usepackage{float}
\usepackage{color}
\usepackage[mathscr]{eucal}
\usepackage{graphics,graphicx,multicol}
\usepackage{epstopdf}
\usepackage{enumerate}
\usepackage{subfigure}
\usepackage{algorithmic}
\usepackage{algorithm}
\usepackage{morefloats}
\usepackage{enumerate}
\usepackage{subfigure}
\usepackage{multirow}
\usepackage{array}
\usepackage{pstricks, pst-node, pst-plot, pst-circ}
\usepackage{moredefs}
\usepackage{courier}
\usepackage{amsbsy}
\usepackage{cite}

\begin{document}
\title{An Application-Aware Spectrum Sharing Approach for Commercial Use of 3.5 GHz Spectrum}
\author{Haya Shajaiah, Ahmed Abdelhadi and Charles Clancy \\
Bradley Department of Electrical and Computer Engineering\\
Hume Center, Virginia Tech, Arlington, VA, 22203, USA\\
\{hayajs, aabdelhadi, tcc\}@vt.edu\\
}
\maketitle
%\thanks{We acknowledge the support from Allied Minds.}
\begin{abstract}
In this paper, we introduce an application-aware spectrum sharing approach for sharing the Federal under-utilized 3.5 GHz spectrum with commercial users. In our model, users are running elastic or inelastic traffic and each application running on the user equipment (UE) is assigned a utility function based on its type. Furthermore, each of the small cells users has a minimum required target utility for its application. In order for users located under the coverage area of the small cells' eNodeBs, with the 3.5 GHz band resources, to meet their minimum required quality of experience (QoE), the network operator makes a decision regarding the need for sharing the macro cell's resources to obtain additional resources. Our objective is to provide each user with a rate that satisfies its application's minimum required utility through spectrum sharing approach and improve the overall QoE in the network. We present an application-aware spectrum sharing algorithm that is based on resource allocation with carrier aggregation to allocate macro cell permanent resources and small cells' leased resources to UEs and allocate each user's application an aggregated rate that can at minimum achieves the application's minimum required utility. Finally, we present simulation results for the performance of the proposed algorithm.
\end{abstract}
% Global Parameters that can be changed:
%%%%%%%%%%%%%%%%%%%%%%%%%%%%%%%%%%%
\providelength{\AxesLineWidth}       \setlength{\AxesLineWidth}{0.5pt}%
\providelength{\plotwidth}           \setlength{\plotwidth}{8cm}% width of the axes only
\providelength{\LineWidth}           \setlength{\LineWidth}{0.7pt}%
\providelength{\MarkerSize}          \setlength{\MarkerSize}{3pt}%
\newrgbcolor{GridColor}{0.8 0.8 0.8}%
\newrgbcolor{GridColor2}{0.5 0.5 0.5}%
%%%%%%%%%%%%%%%%%%%%%%%%%%%%%%%%%%%
%%%%%%%%%%%%%%%%%%%%%%%%%%%%%%%%%%%
\section{Introduction}\label{sec:intro}

The demand for wireless broadband capacity has been recently growing much faster than the availability of new spectrum. Because of the increasing demand for spectrum by commercial wireless operators, federal agencies are now willing to share their spectrum with commercial users. The Commission and the President have outlined a path to double the available spectrum for wireless broadband use, the President’s Council of Advisors on Science and Technology (PCAST) Report identifies two technological advances to increase wireless broadband capabilities. First, increasing the deployment of small cell networks and second using spectrum sharing technology. The 3.5 GHz Band is an ideal band for small cell deployments and shared spectrum use because of its smaller coverage. The National Institute of Standards and Technology (NTIA) Fast Track Report \cite{NTIA10} identified the 3.5 GHz Band for potential shared federal and non-federal broadband use. This band is very favorable for commercial cellular systems such as LTE-Advanced systems.

Small cells are low-powered wireless base stations designed to play well with macro networks in a heterogeneous network (HetNet). Small cells are backed up by a macro cell layer of coverage so that if a small cell shuts down in the 3.5 GHz shared band, operators can pick up coverage again in the macro network.

%%%%The next paragraph is commented to reduce the length of the paper
%The FCC considers finding a market solution for the under-utilized spectrum  through smart regulations. Proposing incentive auctions is one way to do this. In incentive auctions, the licensee gives the spectrum to the highest bidder who either makes revenue and pays a commission to the FCC for running the auction, or shares the revenue with the FCC. Incentive auctions are necessary as they save the high cost and time required for clearing entire bands of spectrum from previous occupants and reallocating them.

Making the under-utilized federal spectrum available for secondary use increases the efficiency of spectrum usage and can provide significant gain in mobile broadband capacity if those resources are aggregated efficiently with the existing commercial mobile systems resources. Many operators are willing to take advantage of the LTE-Advanced carrier aggregation feature which was introduced by 3GPP release 10 \cite{Yuan10CarrierAggregation}. This feature allows users to employ multiple carriers to ensure a wider bandwidth, by aggregating multiple non-continuous or continuous component carriers (CCs), and therefore achieve higher capacity and better performance. In \cite{Haya_Utility1}, the authors have introduced a resource allocation (RA) optimization framework based on carrier aggregation (CA). The proposed multi-stage resource allocation algorithm allocates the primary and secondary carriers resources optimally among users. The final optimal rate allocated to each user is the aggregated rate.% of the primary and secondary carriers. The algorithm uses utility proportional fairness approach to guarantee that no user is allocated zero rate.

In this paper, we introduce an application-aware spectrum sharing approach for cellular networks sharing the federal under-utilized 3.5 GHz spectrum. In our model, the small cells, with the under-utilized 3.5 GHz spectrum resources, are located within the coverage area of a macro cell. The network operator makes a decision regarding the need for sharing the macro cell's eNodeB resources with small cells users based on the small cell users' demand for spectrum resources. %We use utility functions to express the users satisfaction of the service provided to them. Logarithmic utility functions and sigmoidal-like utility functions are used to represent the users delay tolerant applications and real time applications, respectively. In addition,
We use utility proportional fairness approach to guarantee a minimum quality of service (QoS) for each user. In our proposed model, small cells' users have a minimum required utility value for each of their applications. The network operator decides to share the macro cell's eNodeB resources if the value of any of small cell user's application utility function of its allocated rate, i.e. allocated by the small cell's eNodeB, does not exceed the user's application minimum required utility value.
%Only users under the coverage area of both the macro cell and the small cells can benefit from the carrier aggregation approach to achieve at minimum their applications minimum required utility values. We use a convex optimization framework to formulate this resource allocation with CA problem. In our proposed algorithm, the small cell's eNodeB first calculates each of the small cell user's application rate to be allocated to it. If the utility function of any of these rates (i.e. for users under the coverage area of a small cell within the macro cell)  is below the user's application required utility value, the network operator decides to share the macro cell's available resources and starts performing the resource allocation with CA algorithm to find the final aggregated rates and allocate them to each user. In our approach, real-time applications are given priority over delay tolerant applications when allocating the spectrum resources due to their applications nature.

\section{Related Work}\label{Related}
Carrier aggregation enables concurrent utilization of multiple component carriers with different propagation characteristics \cite{LTE-Advanced-Evolving,Dynamic-Spectrum-Refarming}. Due to the significant features of CA, an appropriate CA management is essential to enhance the performance of cellular networks. A tractable multi-band multi-tier CA models for HetNets are proposed in \cite{CA-heterogeneous}. Two models are considered: multi-flow CA and single-flow CA, each UE performs cell selection based on the reference signal's maximum received power. A major concern about deploying small cells is their small coverage areas and low transmit power. The authors in \cite{Heterogeneous-cellular,survey-on-3GPP} have addressed this issue and suggested biasing to allow small cells to expand their coverage areas.

Most of the previous research work have focused on finding resource allocation approaches for intra-system and intra-operator of a single network operator. However, current research on resource allocation are for more complex network topologies \cite{Cognitive-based,Interference-management}. Carrier aggregation in networks that involve multiple network operators in HetNets need to be further investigated. In \cite{methodology-operators}, the authors have analyzed the performance of their proposed carrier aggregation framework that combines a statically assigned spectrum with spectrum resources from a shared spectrum pool.

%The authors in \cite{Dual-Decomposition, Resource_allocation, Rate_Balancing} have studied Resource allocation for single-cell multi-carrier systems. Their  coordinated and collaborative radio resource allocation schemes can significantly improve spectrum utilization. However, it can not be directly deployed in modern networks.
%The RA problem is usually represented as an optimization problem, where the objective is to maximise the overall system throughput that subjects to fairness and transmission power constraints \cite{Fair_resource,Design_of_Fair,Fast_Algorithms,Optimal_and_near-optimal}. Alternatively,
%
The RA optimization problem can be transformed into a utility maximization problem to maximize the user's satisfaction rather than the system throughput, where the user's satisfaction is represented as a function of the achieved data rate \cite{Max-utility-wireless}.
In \cite{Haya_Utility1,Haya_Utility3,Haya_Utility6}, we have proposed a multiple stage RA with CA algorithms that use utility proportional fairness approach to allocate the primary and the secondary carriers resources optimally among mobile users in their coverage area. However, these algorithms consider optimization problems that solve for the allocated rates from the primary and secondary carriers without giving the user or the network operator the flexibility to decide on the amount of recourses to be allocated to the user by secondary carriers. In this paper, we address this issue and design a RA with CA model that accounts for the users' demand of resources and controls which users are required to be allocated additional resources from the secondary carriers. This is important for users who do not wish to pay higher price for more resources if they can be satisfied with certain rates (i.e. rates that guarantee certain degree of satisfaction represented by utility values).
%%%%%%%%%%%%%%%%%%%%%%%%%%%%%%%%%%%%%%%%%%%%%%%%%%%%%%%%%%%%%%%%%%%5
\subsection{Our Contributions}\label{sec:contributions}
Our contributions in this paper are summarized as:
\begin{itemize}
\item We present a spectrum sharing approach for sharing the Federal under-utilized 3.5 GHz spectrum with commercial users.
%\item We prove that the resource allocation optimization problem is convex and therefore the global optimal solution is tractable.
\item We present a spectrum sharing algorithm that is based on resource allocation with CA to allocate the small cells' under-utilized 3.5 GHz resources to small cells' users and allocate the macro cell's resources to both macro cell's users and small cell's users that did not reach their applications minimum required utilities by the small cells allocated rates.
\item We present simulation results for the performance of the proposed resource allocation algorithm.
\end{itemize}

The remainder of this paper is organized as follows. Section \ref{sec:Problem_formulation} presents the problem formulation. In section \ref{sec:RA_optimization}, we present resource allocation optimization problems that solve for the macro cell and small cells allocated rates. Section \ref{sec:Algorithm} presents our proposed resource allocation algorithm. In section \ref{sec:sim}, we discuss simulation setup and provide quantitative results along with discussion. Section \ref{sec:conclude} concludes the paper.

\section{Problem Formulation}\label{sec:Problem_formulation}

We consider LTE-Advanced mobile system consisting of a macro cell, referred to by the index $B$, with a coverage radius $D_B$, that is overlaid with $S$ small cells. The macro cell's eNodeB is configured at the LTE-Advanced carrier and the small cell's eNodeB is configured to use the 3.5 GHz under-utilized spectrum band.
Let $\mathcal{S}$ denotes the set of small cells located within the coverage area of the macro cell $B$ where $S=|\mathcal{S}|$. All small cells are connected to the core network. The small cells are assumed to have a closed access scheme where only registered UEs, referred to by SUEs, are served by the small cells eNodeBs. On the other hand, all UEs under the coverage area of the macro cell $B$ and not within the coverage of any small cell, referred to by MUEs, are served by the macro cell's eNodeB. The set of all MUEs under the coverage area of macro cell $B$ is referred to by $\mu$. The set of SUEs associated to small cell $s$ is referred to by $\mathcal{Q}_s$. We assume that the association of the UEs with their eNodeBs remains fixed during the runtime of the resource allocation process. We have $\bigcup_{s=1}^{S}\mathcal{Q}_s=\Theta$ and $\bigcap_{s=1}^{S}\mathcal{Q}_s=\emptyset$.
Each SUE $i$ has a minimum QoE requirement for its applications that is represented by the utility of the user's application with its allocated rate. Let $u_i^{\text{req}}$ denotes the minimum required utility of SUE $i \in \Theta$.

Utility functions are used to express the user satisfaction with its allocated rate \cite{DL_PowerAllocation,Fundamental,Utility-proportional,Ahmed_Utility2}. The $i^{th}$ user application utility function of its allocated rate $r_i$ is given by $U_i(r_i)$ where $U_i$ is a sigmoidal-like function used to represent real time applications or logarithmic function used to represent delay tolerant applications. These utility functions have the following properties:
\begin{itemize}
\item $U_i(0) = 0$ and $U_i(r_i)$ is an increasing function of $r_i$.
\item $U_i(r_i)$ is twice continuously differentiable in $r_i$ and bounded above.
\end{itemize}

In our model, we use normalized sigmoidal-like utility functions, as in \cite{Ahmed_Utility1}, that are expressed as
\begin{equation}\label{eqn:sigmoid}
U_i(r_i) = c_i\Big(\frac{1}{1+e^{-a_i(r_i-b_i)}}-d_i\Big),
\end{equation}
where $c_i = \frac{1+e^{a_ib_i}}{e^{a_ib_i}}$ and $d_i = \frac{1}{1+e^{a_ib_i}}$ so it satisfies $U_i(0)=0$ and $U_i(\infty)=1$. The normalized sigmoidal-like function has an inflection point at $r_i^{\text{inf}}=b_i$. In addition, we use the normalized logarithmic utility function, used in \cite{Ahmed_Utility1}, that are expressed as
%The normalized sigmoidal-like utility functions with $a=5$ and $b=10$, and $a=5$ and $b=10$ is shown in Figure \ref{fig:sigmoid}.
\begin{equation}\label{eqn:log}
U_i(r_i) = \frac{\log(1+k_ir_i)}{\log(1+k_ir_i^{\text{max}})},
\end{equation}
where $r_i^{\text{max}}$ gives $100\%$ utilization and $k_i$ is the rate of increase of utility percentage with allocated rates that varies based on the user application. So, it satisfies $U_i(0)=0$ and $U_i(r_i^{\text{max}})=1$.

\begin{figure}[tb]
\centering
\includegraphics[width=3.5in]{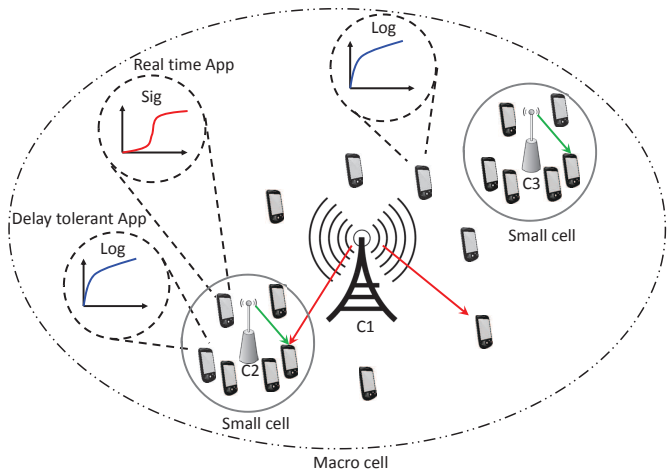}
\caption{System model for a LTE-Advanced mobile system with one macro cell and two small cells within the coverage area of the macro cell. Each of the small cells is configured to use the 3.5 GHz under-utilized spectrum.}
\label{fig:SystemModel}
\end{figure}

%%%%%%%%%%%%%%%%%%%%%%%%%%%%%%%%%%%%%%%%
Figure \ref{fig:SystemModel} shows a heterogeneous network that consists of one macro cell with one eNodeB and two small cells within the coverage area of the macro cell, each of the small cells has one eNodeB that is configured to use the 3.5 GHz under-utilized spectrum. Mobile users under the coverage of the macro cell and the small cells are running real time or delay tolerant applications that are represented by sigmoidal-like or logarithmic utility functions, respectively.
%%%%%%%%%%%%%%%%%%%%%%%%%%%%%%%%%%%%%%%%

\section{Resource Allocation Optimization for Spectrum Sharing with the 3.5 GHz Spectrum}\label{sec:RA_optimization}
In this section, we present a resource allocation framework for cellular networks sharing the federal under-utilized 3.5 GHz spectrum. In our model, SUEs are allocated resources from the leased under-utilized 3.5 GHz resources at the small cells eNodeBs whereas MUEs are allocated resources only by the macro cell's eNodeB. Each of the SUEs has a minimum required utility $u_i^{\text{req}}$ for each of its applications. First the small cell's eNodeB allocates its available leased resources then the network operator decides which SUEs still require additional resources in order to achieve their minimum required utilities and allocate them more resources from the macro cell eNodeB based on a resource allocation with carrier aggregation optimization problem.

The resource allocation process starts by allocating each of the small cells resources to SUEs under it coverage area. We use a utility proportional fairness resource allocation optimization problem to allocate the small cell resources. The RA optimization problem of the small cell $s$ is given by:
\begin{equation}\label{eqn:opt_SmallCell}
\begin{aligned}
& \underset{\textbf{r}^s}{\text{max}}
& & \prod_{i=1}^{|\mathcal{Q}_s|}U_i(r_i^s) \\
& \text{subject to}
& & \sum_{i=1}^{|\mathcal{Q}_s|}r_i^s \leq R_s\\
& & &  0 \leq r_i^s \leq R_s, \;\;\;\;\; i = 1,2, ...,|\mathcal{Q}_s|,
\end{aligned}
\end{equation}
where $\mathbf{r}^s =\{r_{1}^s,r_{2}^s,...,r_{|\mathcal{Q}_s|}^s\}$, $|\mathcal{Q}_s|$ is the number of SUEs under the coverage area of the small cell $s$ and $R_s$ is the maximum achievable rate of the under-utilized 3.5 GHz leased spectrum available at the eNodeB of small cell $s$. The resource allocation objective function is to maximize the entire small cell utility when allocating its resources. It also achieves proportional fairness among utilities such that non of the SUEs will be allocated zero resources. Therefore, a minimum QoS is provided to each SUE. This approach gives real time applications priority when allocating the small cell resources. The objective function in optimization problem (\ref{eqn:opt_SmallCell}) is equivalent to $\underset{\mathbf{r}^s}{\text{max}} \sum_{i=1}^{|\mathcal{Q}_s|} \log U_i(r_{i}^s)$. Optimization problem (\ref{eqn:opt_SmallCell}) is a convex optimization problem and there exists a unique tractable global optimal solution \cite{Ahmed_Utility1}.

From optimization problem (\ref{eqn:opt_SmallCell}), we have the Lagrangian:
\begin{equation}\label{eqn:Lagrangian SmallCell}
\begin{aligned}
L_s(\textbf{r}^s,p^s) = (\sum_{i=1}^{|\mathcal{Q}_s|}\log U_i(r_{i}^s))
-p^s(\sum_{i=1}^{|\mathcal{Q}_s|}r_{i}^s+z_s-R_s)
\end{aligned}
\end{equation}
where $z_s\geq 0$ is the slack variable and $p^s$ is the Lagrange multiplier which is equivalent to the shadow price that corresponds to the service provider's price per unit bandwidth for the small cell resources \cite{Ahmed_Utility1}.

The solution of equation (\ref{eqn:opt_SmallCell}) is given by the values $r_i^s$ that solve equation $\frac{\partial \log U_i(r_i^s)}{\partial r_i^s} = p^s$ and are the intersection of the time varying shadow price, horizontal line $y = p^s$, with the curve $y = \frac{\partial \log U_i(r_i^s)}{\partial r_i^s}$ geometrically. Once the RA process is performed by the small cell $s$, each SUE in $\mathcal{Q}_s$ will be allocated $r_i^{s,\text{all}}=r_i^s$ rate. However, the network operator decides if any of the SUEs requires additional resources in order to reach the minimum required utility $u_i^{\text{req}}$ of its application by comparing the utility of the small cell allocated rate that is given by $U_i(r_i^{s,\text{all}})$ with the value $u_i^{\text{req}}$. If the achieved utility for certain SUE is less that the minimum required utility, the network operator requests additional resources from the macro cell for that SUE. The small cell $s$ eNodeB creates a set $\mathcal{Q}_{sB}$ of all SUEs that needs to be allocated additional resources where $\mathcal{Q}_{sB}=\{\text{SUEs} \in \mathcal{Q}_s \;\; s.t. \;\; u_i^{\text{req}} > U_i(r_i^{s,\text{all}})\}$.

Once each small cell $s$ within the coverage area of the macro cell $B$ performs its RA process based on optimization problem (\ref{eqn:opt_SmallCell}), the macro cell starts allocating its resources to all MUEs within its coverage area as well as the SUEs that were reported, by the network operator, for their need of additional resources. Let $\mathcal{Q}$ be the set of SUEs that will be allocated additional resources by the macro cell where $\mathcal{Q}=\bigcup_{s=1}^{S} \mathcal{Q}_{sB}$. The set of UEs that will be served by the macro cell's eNodeB; i.e. participate in the macro cell RA process, is given by $\beta$ where $\beta=\mu \bigcup \mathcal{Q}$. The resource allocation optimization problem of the macro cell $B$ is given by:

\begin{equation}\label{eqn:opt_MacroCell}
\begin{aligned}
& \underset{\textbf{r}}{\text{max}}
& & \prod_{i=1}^{|\beta|}U_i(r_i+C_i) \\
& \text{subject to}
& & \sum_{i=1}^{|\beta|}r_i \leq R_B\\
& & & C_i=
\begin{cases}
	0 \;\;\;\;\;\;\; \text{if UE}\;\; i \notin \mathcal{Q}\\
	r_i^{s,\text{all}} \;\;\text{if UE}\;\; i \in \mathcal{Q}
\end{cases}\\
& & &  0 \leq r_i \leq R_B, \;\;\;\;\; i = 1,2, ...,|\beta|,
\end{aligned}
\end{equation}
where $\mathbf{r} =\{r_{1},r_{2},...,r_{|\beta|}\}$, $|\beta|$ is the number of UEs that will be be served by the macro cell's eNodeB and $R_B$ is the maximum achievable rate of the resources available at the macro cell's eNodeB. The resource allocation objective function is to maximize the entire macro cell utility when allocating its resources. The RA optimization problem (\ref{eqn:opt_MacroCell}) is based on carrier aggregation. It seeks to maximize the multiplication of the utilities of the rates allocated to MUEs by the macro cell's eNodeB and the utilities of the rates allocated to the SUEs in $\beta$ by small cells' eNodeBs and macro cell's eNodeB. Utility proportional fairness is used to guarantee that non of the UEs will be allocated zero resources. Real time applications are given priority when allocating the macro resources using this approach. The objective function in optimization problem (\ref{eqn:opt_MacroCell}) is equivalent to $\underset{\mathbf{r}}{\text{max}} \sum_{i=1}^{|\beta|} \log U_i(r_i+C_i)$. Optimization problem (\ref{eqn:opt_MacroCell}) is a convex optimization problem and there exists a unique tractable global optimal solution \cite{Ahmed_Utility1}.

From optimization problem (\ref{eqn:opt_MacroCell}), we have the Lagrangian:
\begin{equation}\label{eqn:Lagrangian MacroCell}
\begin{aligned}
L_B(\textbf{r},p^B) = (\sum_{i=1}^{|\beta|}\log U_i(r_{i}+C_i))
-p^B(\sum_{i=1}^{|\beta|}r_{i}+z_B-R_B)
\end{aligned}
\end{equation}
where $z_B\geq 0$ is the slack variable and $p^B$ is the Lagrange multiplier which is equivalent to the shadow price that corresponds to the service provider's price per unit bandwidth for the macro cell resources \cite{Ahmed_Utility1}.

The solution of equation (\ref{eqn:opt_MacroCell}) is given by the values $r_i$ that solve equation $\frac{\partial \log U_i(r_i+C_i)}{\partial r_i} = p^B$ and are the intersection of the time varying shadow price, horizontal line $y = p^B$, with the curve $y = \frac{\partial \log U_i(r_i+C_i)}{\partial r_i}$ geometrically. Once the macro cell eNodeB is done performing the RA process based on optimization problem (\ref{eqn:opt_MacroCell}), each UE in $\beta$ will be allocated $r_i^{\text{all}}=r_i+C_i$ rate.

\section{The Macro Cell and Small Cells RA Optimization Algorithm}\label{sec:Algorithm}

In this section, we present our resource allocation algorithm. The proposed algorithm consists of SUE, MUE, small cell eNodeB and macro cell eNodeB parts shown in Algorithm \ref{alg:SUE}, \ref{alg:MUE}, \ref{alg:s_eNodeB} and \ref{alg:B_eNodeB}, respectively. The execution of the algorithm starts by SUEs and MUEs, subscribing for mobile services, transmitting their applications utilities parameters to their corresponding eNodeBs. First, each small cell $s$ eNodeB calculates its allocated rate $r_i^{s,\text{all}}$ to each SUE in $\mathcal{Q}_s$. It then checks whether the achievable utility of that rate is less or greater than the SUE's minimum required utility $u_i^{\text{req}}$. If for any SUE $U_i(r_i^{s,\text{all}}) < u_i^{\text{req}}$, the small cell's eNodeB sends the application parameters and the allocated rate $r_i^{s,\text{all}}$  for that SUE to the macro cell's eNodeB requesting additional resources. Otherwise, it allocates the rate $r_i^{s,\text{all}}$ to that SUE.

Once the macro cell's eNodeB receives the set $\mathcal{Q}_{sB}$ from each small cell in $\mathcal{S}$ within its coverage area. It starts the RA process to allocate its available resources to each UE in $\beta$ based on a RA with carrier aggregation optimization problem. Once the RA process of the macro cell is performed, the macro cell allocates rate $r_i^{\text{all}}=r_i+C_i$ to the $i^{th}$ UE in $\beta$.

\begin{algorithm}
\caption{The $i^{th}$ SUE $\in \mathcal{Q}_s$ Algorithm}\label{alg:SUE}
\begin{algorithmic}
\LOOP
      \STATE {Send application utility parameters $k_i$, $a_i$, $b_i$, $r_i^{\text{max}}$ and $u_i^{\text{req}}$ to the SUE's in band small cell's eNodeB.}
      \STATE {Receive the final allocated rate $r_i^{s,\text{all}}$ from the small cell $s$ eNodeB or from the macro cell's eNodeB.}
\ENDLOOP
\end{algorithmic}
\end{algorithm}

\begin{algorithm}
\caption{The $i^{th}$ MUE $\in \mu$ Algorithm}\label{alg:MUE}
\begin{algorithmic}
\LOOP
      \STATE {Send application utility parameters $k_i$, $a_i$, $b_i$ and $r_i^{\text{max}}$ to the macro cell's eNodeB.}
      \STATE {Receive the final allocated rate $r_i^{\text{all}}$ from the macro cell's eNodeB.}
\ENDLOOP
\end{algorithmic}
\end{algorithm}

%%%%eNodeB pseudocode
\begin{algorithm}
\caption{Small Cell $s$ eNodeB Algorithm}\label{alg:s_eNodeB}
\begin{algorithmic}
\LOOP
\STATE {Initialize $\mathcal{Q}_{sB}=\emptyset$; $r_i^{\text{all}}=0$.}
\STATE {Receive application utility parameters $k_i$, $a_i$, $b_i$, $r_i^{\text{max}}$ and $u_i^{\text{req}}$ from all SUEs in $\mathcal{Q}_s$.}%
 \STATE{Solve $\textbf{r}^s =  \arg \underset{\textbf{r}^s}\max \sum_{i=1}^{|\mathcal{Q}_s|}\log U_i(r_i^s) - p^s(\sum_{i=1}^{|\mathcal{Q}_s|}(r_i^s)-R_s)$.}
 \STATE{Let $r_i^{s,\text{all}}=r_i^s$ be the rate allocated by the $s$ small cell's eNodeB to each user in $\mathcal{Q}_s$.}
 \STATE{Calculate the SUE utility  $U_i(r_i^{s,\text{all}}) \:\:\forall i \in \mathcal{Q}_s$}
\FOR {SUE $i \leftarrow 1$  to  $|\mathcal{Q}_s|$}
 \IF{$U_i(r_i^{s,\text{all}}) < u_i^{\text{req}}$}
 \STATE{$\mathcal{Q}_{sB}=\mathcal{Q}_{sB} \bigcup$ SUE$\{i\}$}
 \STATE{Send SUE $i$ parameters $k_i$, $a_i$, $b_i$, $r_i^{\text{max}}$ and $r_i^{s,\text{all}}$ to the macro cell's eNodeB}
 \ELSE
 \STATE{Allocate rate $r_i^{\text{all}}=r_i^{s,\text{all}}$ to SUE $i$}
 \ENDIF
\ENDFOR
\ENDLOOP
\end{algorithmic}
\end{algorithm}

%%%%eNodeB pseudocode
\begin{algorithm}
\caption{The Macro Cell's eNodeB Algorithm}\label{alg:B_eNodeB}
\begin{algorithmic}
\LOOP
\STATE {Initialize $C_i=0$; $r_i^{\text{all}}=0$.}
\FOR {$s \leftarrow 1$ to  $S$}
\STATE {Receive application utility parameters $k_i$, $a_i$, $b_i$, $r_i^{\text{max}}$ and $r_i^{s,\text{all}}$ for all SUEs in $\mathcal{Q}_{sB}$ from small cell $s$ eNodeB.}%
\STATE{$C_i=r_i^{s,\text{all}} \:\:\forall i \in \mathcal{Q}_{sB}$}
\ENDFOR
\STATE {Create user group $\mathcal{Q}=\bigcup_{s=1}^{S} \mathcal{Q}_{sB}$}
\STATE {Create user group $\beta=\mu \bigcup \mathcal{Q}$}
 \STATE{Solve $\textbf{r} =  \arg \underset{\textbf{r}}\max \sum_{i=1}^{|\beta|}\log U_i(r_i+C_i) - p^B(\sum_{i=1}^{|\beta|}(r_i)-R_B)$.}
 \STATE{Allocate $r_i^{\text{all}}=r_i+C_i$ to each UE $i$ in $\beta$}
\ENDLOOP
\end{algorithmic}
\end{algorithm}

\section{Simulation Results}\label{sec:sim}
\begin{figure}[tb]
\centering
\includegraphics[height=1.9in,width=3.5in]{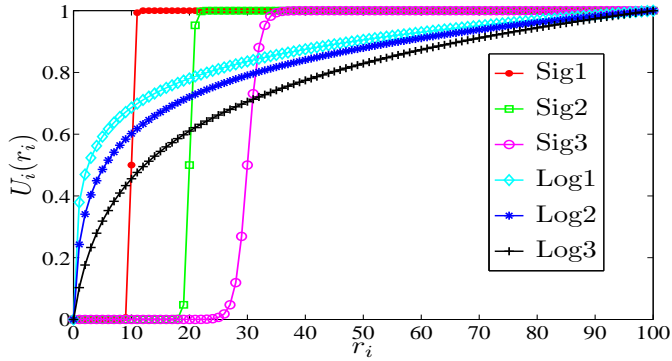}
\caption{The users utility functions $U_i(r_i)$ used in the simulation (three sigmoidal-like functions and three logarithmic functions).}
%%\myfigureshrinker{\vspace{-0.06in}}
\label{fig:utility}
\end{figure}

Algorithm \ref{alg:SUE}, \ref{alg:MUE}, \ref{alg:s_eNodeB} and \ref{alg:B_eNodeB} were applied in C++ to multiple utility functions with different parameters. Simulation results showed convergence to the global optimal rates. In this section, we consider a macro cell with one eNodeB. Within the the coverage area of the macro cell there exists one small cell $s$. Four SUEs are located under the coverage area of the small cell $s$ with UEs indexes $\{1,2,3,4\}$. The SUEs user group is given by $\mathcal{Q}_s=\{1,2,3,4\}$. Four MUEs are located under the coverage area of the macro cell's eNodeB but not within the small cell. The MUEs user group is given by $\mu=\{5,6,7,8\}$. Each UE whether it is SUE or MUE is running either real time application or delay tolerant application. Each of the SUEs' applications utilities has a minimum required utility that is given by $u_i^{\text{req}}$ that is equivalent to the $C_i$ value for that user whereas MUEs do not have minimum required utilities for their applications. The UEs' indexes, types and applications utilities parameters are listed in table \ref{table:parameters}. Figure \ref{fig:utility} shows the sigmoidal-like utility functions and the logarithmic utility functions used to represent the SUEs and MUEs applications.

\begin {table}[]
\caption {Users and their applications utilities}
\label{table:parameters}
\begin{center}
\renewcommand{\arraystretch}{1.4} %<- modify value to suit your needs
\begin{tabular}{| l | l | l | }
%\label{table:utility}
  \hline
  \multicolumn{1}{|c|}{User's Index} & \multicolumn{1}{|c|}{User's Type} & \multicolumn{1}{|c|}{Applications Utilities Parameters} \\  \hline
  UE1 $i=\{1\}$ & SUE & Sig2 $a_i=3,\: b_i=20,\: u_i^{\text{req}}=0.8$\\ \hline
  UE2 $i=\{2\}$ & SUE & Sig3 $a_i=1,\: b_i=30,\: u_i^{\text{req}}=0.8$\\ \hline
  UE3 $i=\{3\}$ & SUE & Log2 $k_i=3,\: r_i^{\text{max}}=100,\: u_i^{\text{req}}=0.5$\\ \hline
  UE4 $i=\{4\}$ & SUE & Log3 $k_i=0.5,\: r_i^{\text{max}}=100,\: u_i^{\text{req}}=0.5$\\ \hline
  UE5 $i=\{5\}$ & MUE & Sig1 $a_i=5,\: b_i=10$\\ \hline
  UE6 $i=\{6\}$ & MUE & Sig3 $a_i=1,\: b_i=30$\\ \hline
  UE7 $i=\{7\}$ & MUE & Log1 $k_i=15,\: r_i^{\text{max}}=100$\\ \hline
  UE8 $i=\{8\}$ & MUE & Log3 $k_i=0.5,\: r_i^{\text{max}}=100$\\ \hline

\end{tabular}
%\caption {Should be a caption}
\end{center}
\end {table}
%%%%%%%%%%%%%%%%%%%%%%%%%%%%%%%%%%%%%%%%%%%%%%
%%%%%%%%%%%%%%%%%%%%%%%%%
%%%%%%%%%%%%%%%%%%%%%%%%%%%%%%
%%%%%%%%%%%%%%%%%%%%%%%%%%%%%%%%%%%%%%%%%%%%%%
\subsection{Small Cell Allocated Rates and Users QoE}\label{subsec:sim_SmallCell}
\begin{figure}[tb]
  \centering
  \subfigure[The rates $r_i^{s,\text{all}}$ allocated by the small cell's eNodeB to users in $\mathcal{Q}_s$ with $10<R_s<100$.]{%
  \label{fig:ri_SmallCell_R_s}
  \includegraphics[height=1.9in,width=3.5in]{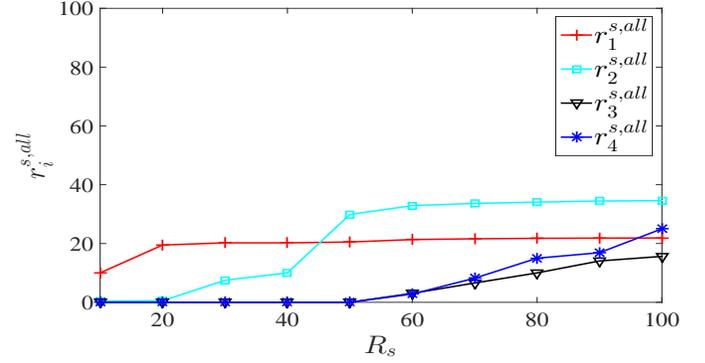}
  }\\%
\subfigure[Users' QoE represented by the utility of user's application of its allocated rate $U_i(r_i^{s,\text{all}})$ when $R_s=50$ and $R_s=70$.]{%
  \label{fig:UtilitySmallCell}
  \includegraphics[height=1.7in,width=3.5in]{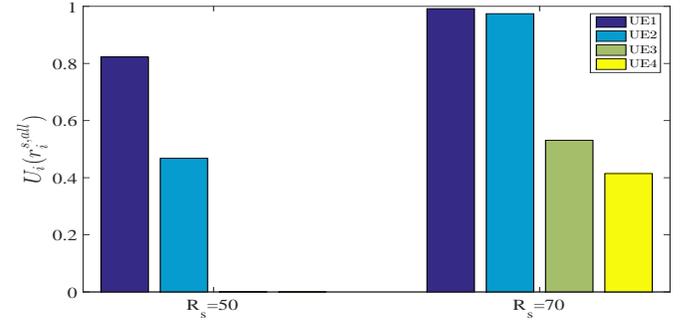}
  }%
\caption{The small cell's eNodeB allocated rates with $10<R_s<100$ and users' QoE when $R_s=50$ and $R_s=70$.}
\label{fig:SmallCell}
\end{figure}

In the following simulations, the small cell's carrier total rate $R_s$ takes values between $10$ and $100$ with step of $10$. In Figure \ref{fig:SmallCell}, we show the small cell's allocated rates $r_i^{s,\text{all}}$ for users in $\mathcal{Q}_s$ with different values of the small cell's carrier total rate $R_s$ and the users QoE with the small cell allocated rates when $R_s=50$ and $R_s=70$. In Figure \ref{fig:ri_SmallCell_R_s}, we show that users running real time applications are given priority when allocating the small cell's resources due to their sigmoidal-like utility function nature. We also observe that non of the UEs is allocated zero resources because we used a utility proportional fairness approach. We also show how the proposed rate allocation algorithm converges for different values of $R_s$. In Figure \ref{fig:UtilitySmallCell}, we show the QoE for the four SUEs which is represented by their applications utilities of the small cell allocated rates $U_i(r_i^{s,\text{all}})$ when $R_s=50$ and $R_s=70$. We notice that in the case of $R_s=50$, the utilities of the small cell allocated rates for UE2, UE3 and UE4 did not reach the minimum required utilities for these SUEs whereas in the case of $R_s=70$ the utility of the small cell allocated rate for UE4 did not reach the minimum required utility for that SUE. Therefore, based on the proposed algorithm the network operator will request additional resources for these UEs from the macro cell's eNodeB and these UEs will be allocated additional resources based on a resource allocation with carrier aggregation scenario.
%%%%%%%%%%%%%%%%%%%%%%%%%%%%%%%%%%%%%%%%%%%%%%%%%
\subsection{Macro Cell Allocated Rates and Users QoE}
\begin{figure}[tb]
  \centering
  \subfigure[The aggregated rates $r_i^{\text{all}}=r_i+C_i$ allocated by the macro cell's eNodeB to users in $\beta$ when $R_s=50$.]{%
  \label{fig:ri_MacroCell_R_B}
  \includegraphics[height=1.9in,width=3.5in]{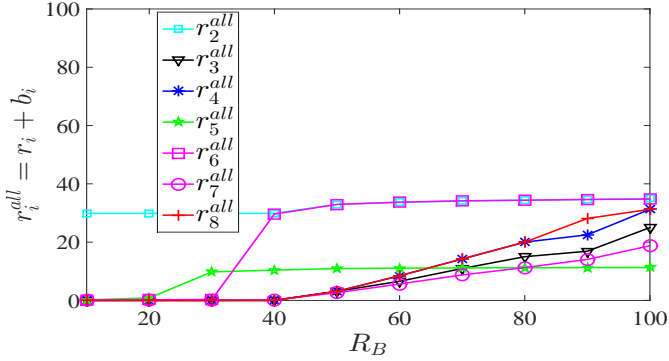}
  }\\%
\subfigure[Users' QoE represented by the utility of user's application of its allocated rate $U_i(r_i^{\text{all}})$ when $R_B=80$ and $R_s=50$.]{%
  \label{fig:UtilityMacroCell}
  \includegraphics[height=1.6in,width=3.5in]{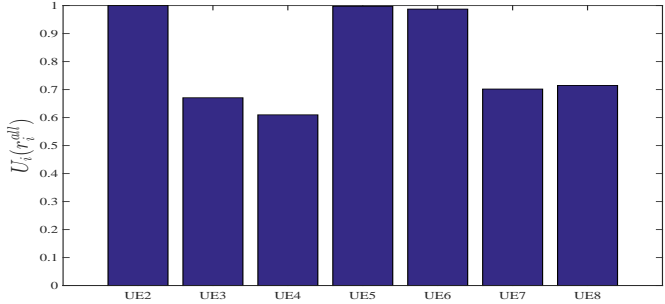}
  }%
\caption{The total aggregated rates $r_i^{\text{all}}=r_i+C_i$ allocated by the macro cell's eNodeB to users in $\beta$ with $10<R_B<100$ when $R_s=50$ and the users' QoE when $R_B=80$ and $R_s=50$.}
\label{fig:MacroCell}
\end{figure}

In the following simulations, the macro cell's carrier total rate $R_B$ takes values between $10$ and $100$ with step of $10$ and $R_s$ is fixed at $50$. As discussed in \ref{subsec:sim_SmallCell}, in the case of $R_s=50$ the network operator requests additional resources for three SUEs (i.e. UEs in $\mathcal{Q}_{sB}=\{2,3,4\}$) as they did not reach their minimum required utilities. Therefore, the macro cell's eNodeB performs a resource allocation with carrier aggregation process to allocate resources to the UEs in user group $\beta$ where $\beta=\{2,3,4,5,6,7,8\}$. In Figure \ref{fig:MacroCell}, we show the final allocated rates $r_i^{\text{all}}$ for the UEs in $\beta$ and these users QoE with the final allocated rates when $R_B=80$. In Figure \ref{fig:ri_MacroCell_R_B}, we show the macro cell's final allocated rates converges for different values of $R_B$. Again we observe that non of the users is allocated zero resources and that real time applications are given priority when allocating the macro cell's resources. In Figure \ref{fig:UtilityMacroCell}, we show the QoE for the seven UEs in $\beta$ which is represented by their applications utilities of the final allocated rate $U_i(r_i^{\text{all}})$ when $R_s=50$ and $R_B=80$. We notice that the utilities of the final allocated rates for the three SUEs in $\mathcal{Q}_{sB}$ (i.e. UE \{2,3,4\}) exceed the minimum required utilities for these SUEs because of the additional resources allocated to these users by the macro cell's eNodeB.
%%%%%%%%%%%%%%%%%%%%%%%%%%%%%%%%%%%%%%%%%%%%%%%%%

\section{Conclusion}\label{sec:conclude}
In this paper, we proposed a spectrum sharing approach for sharing the Federal under-utilized 3.5 GHz spectrum with commercial users. We used sigmoidal-like utility functions and logarithmic utility functions to represent real time and delay tolerant applications, respectively. We presented resource allocation optimization problems that are based on carrier aggregation. The proposed resource allocation algorithm ensures fairness in the utility percentage. Users located under the coverage area of the small cells are allocated resources by the small cells' eNodeBs whereas both the macro cell users and the small cells' users that did not reach their minimum required utilities by their small cells' allocated rates are allocated resources by the macro cell's eNodeB based on carrier aggregation. We showed through simulations the the proposed algorithm converges to the optimal rates. We also showed that small cells' users can achieve their minimum required QoE by using the proposed spectrum sharing approach.

\bibliographystyle{ieeetr}
\bibliography{pubs}
\end{document}